\documentclass[aps,prl,twocolumn,intlimits,showpacs]{revtex4-1}
\usepackage{newtxtext,newtxmath}
\usepackage{amsmath,amssymb}
\usepackage{graphicx}
\usepackage[T1]{fontenc}
\usepackage{tikz}
\usepackage[hypcap]{caption}
\usepackage{subcaption}
\usepackage{bbm}
\usepackage{url,parskip}
\usepackage{hyperref}
\usepackage[english]{babel}

\DeclareSymbolFont{largesymbolsCM}{OMX}{cmex}{m}{n}

\let\sum\relax
\DeclareMathSymbol{\sum}{\mathop}{largesymbolsCM}{"50}

\usepackage[compact]{titlesec}	
\titlespacing\section{0pt}{0pt}{0pt}
\hypersetup{
	bookmarks=true,         
	unicode=false,          
	pdftoolbar=true,        
	pdfmenubar=true,        
	pdffitwindow=false,    
	pdfstartview={FitH},    
	pdftitle={My title},    
	pdfauthor={Author},     
	pdfsubject={Subject},   
	pdfcreator={Creator},   
	pdfproducer={Producer}, 
	pdfkeywords={keyword1} {key2} {key3}, 
	pdfnewwindow=true,      
	colorlinks=true,       
	linkcolor=blue,          
	citecolor=blue,      
	filecolor=green,       
	urlcolor= black,         
}

\begin{document}
\title{The one-dimensional Kardar-Parisi-Zhang and Kuramoto-Sivashinsky 
universality class: limit distributions}
\author{Dipankar Roy}\email{dipankarroy@iisc.ac.in}
\affiliation{Department of Mathematics, Indian Institute of Science,\\
	Bangalore - 560012, India.}
\author{Rahul Pandit}\email{rahul@iisc.ac.in ; also at Jawaharlal Nehru Centre 
	for Advanced Scientific Research, Jakkur, Bangalore 560 064} 
\affiliation{Centre for Condensed Matter Theory, Department of Physics,
	Indian Institute of Science,\\
	Bangalore - 560012, India.}

\date{\today}
\begin{abstract}

Tracy-Widom and Baik-Rains distributions appear as universal limit
distributions for height fluctuations in the one-dimensional
Kardar-Parisi-Zhang (KPZ) \textit{stochastic} partial differential
equation (PDE).  We obtain the same universal distributions in the
spatiotemporally chaotic, nonequilibrium, but statistically steady
state (NESS) of the one-dimensional Kuramoto-Sivashinsky (KS)
\textit{deterministic} PDE, by carrying out extensive pseudospectral 
direct numerical
simulations to obtain the spatiotemporal evolution of the KS height
profile $h(x,t)$ for different initial conditions. We establish,
therefore, that the statistical properties of the 1D KS PDE in this
state are in the 1D KPZ universality class.

\end{abstract}

\pacs{02.30.Jr,05.10.-a,47.70.-n,68.35.Rh,74.40.Gh}

\keywords{Kuramoto-Sivashinsky equation, Kardar-Parisi-Zhang equation, 
Tracy-Widom distribution, Baik-Rains distribution.}

\maketitle

Fundamental investigations of the statistical properties of hydrodynamical
turbulence often use \textit{randomly forced} versions of the
\textit{deterministic} Navier-Stokes (NS) equations (3D NSE, in three
dimensions); the latter use a non-random forcing term to produce a turbulent,
but nonequilibrium, statistically steady state (NESS). A randomly forced 3D,
incompressible NS equation (3D RFNSE), proposed first by
Edwards~\cite{edwards1964} in 1964, has been studied extensively, via
renormalization-group (RG) and other theoretical~\cite{fns1977, dm1979, ff1983,
yo1986, mw1995, jkb1988, aav1996, aav1999} and numerical~\cite{smp1998,
bclst2004} methods; these studies have shown that many statistical properties
of turbulence in the 3D RFNSE are akin to their 3D NSE counterparts. In
particular, the wave-number $k$ dependence of the energy spectrum~\cite{k1941a,
k1941b, f1995} $E(k)$, and even the mutiscaling corrections~\cite{f1995,
pf1985, bppv1984, bf2010, ms1991} to the Kolmogorov phenomenology~\cite{k1941a,
k1941b, f1995} of 1941 are similar in both these models.

Can we find such similarity between the statistical properties of NESSs in
\textit{deterministic} and related \textit{stochastic} partial differential
equations (PDEs) that are simpler than their 3D hydrodynamical counterparts? It
has been suggested, since the 1980s, that the Kuramoto-Sivashinsky (KS) PDE, a
deterministic interface-growth model for a height field $h({\bf x},t)$, which
is used in studies of chemical waves, flame fronts, and the surfaces of thin
films flowing under gravity~\cite{kuramoto1976,siva1977, sm1980, ruyer1998,
pz1985, chen1984, grinstein1996}, is a simplified model for
turbulence~\cite{pz1985}. It has been conjectured~\cite{Yakhot1981}, and
subsequently shown by compelling numerical studies~\cite{hnz1986, sneppen1992,
hayot1993, jayaprakash1993, 2d_bch1999, 2d_kkp2015}, in both one dimension (1D)
and two dimensions (2D), that the long-distance and long-time behaviors of
correlation functions, in the spatiotemporally chaotic NESS of the KS PDE,
exhibit the same power-law scaling as their couterparts in the the
Kardar-Parisi-Zhang (KPZ)
equation~\cite{kpz1986,thhzhang1995,thhkat2015,quastel2015}, a stochastic PDE
(SPDE), in which the height field $h({\bf x},t)$ is kinetically roughened.  The
elucidation of the statistics of $h({\bf x},t)$ in the KPZ SPDE has played a
central role in nonequilibrium statistical mechanics, in general, and
interface-growth phenomena, in particular.  Early KPZ
studies~\cite{kpz1986,thhzhang1995} have concentrated on height-field
correlations, the width $w(L,t)$ of the fluctuating KPZ interface, and their
power-law dependences on the linear system size $L$ and time $t$, for large $L$
and $t$ (see below); especially for the 1D case, several results can be
obtained analytically. The universality of the power-law exponents has been
demonstrated by explicit numerical calculations, e.g., in the poly-nuclear
growth (PNG) model, directed polymers in random media (DPRM), and the
asymmetric simple exclusion process (ASEP), and by experiments in turbulent
liquid crystals~\cite{takeuchi2011,takeuchi2012,takeuchi2013}, all of which lie
(in suitable parameter regimes) in the KPZ universality class.  The seminal
work of Pr\"{a}hofer and Spohn work (recently referred to as ``the $2^{nd}$ KPZ
Revolution''~\cite{thhkat2015})  on the PNG model~\cite{prahofer2000} has led
to a new set of studies of the 1D KPZ universality
class~\cite{sasamoto2010,calabrese2011,imamura2012,corwin2012,thhlin2014,quastel2015,saberi-naserabadi-krug-2019},
which have led to the remarkable result that, at a point $x$ and at large times
$t$,
\begin{equation}
	h(x,t) - h(x,0) \approx  v_{\infty} t + ( \Gamma t)^{\upbeta_{\text{KPZ}}} \upchi_\beta + 
	o(t^{\upbeta_{\text{KPZ}}}) \ , \ \text{for} \ t \rightarrow \infty,
	\label{eq:KPZh}
\end{equation}
where $v_{\infty}$ and $\Gamma$ are model-dependent constants (Supplemental
Material~\cite{supp}), the exponent $\upbeta_{\text{KPZ}}=1/3$, and
$\upchi_\beta$ is a random variable distributed according to the Tracy-Widom
(TW) distribution for the Gaussian Orthogonal Ensemble (GOE) ($\beta=1$) and
for the Gaussian Unitary Ensemble (GUE) ($\beta=2$), familiar from the theory
of random matrices~\cite{tracy1994}, or the Baik-Rains (BR $F_{0}$)
distribution~\cite{baik2000} ($\beta=0$); the value of $\beta$ depends on the
initial condition. We show, by extensive direct numerical simulations (DNSs),
that the result~\eqref{eq:KPZh} holds for the NESS of the 1D KS PDE. Thus, the
correspondence between the statistical properties of these states, in the  1D
KS (PDE) and their counterparts in the 1D KPZ (SPDE), does not stop at the
simple correlation functions, investigated so far~\cite{hnz1986, sneppen1992,
hayot1993, jayaprakash1993}; we demonstrate that this correspondence includes
the universal limit distributions obtained in ``the $2^{nd}$ KPZ
Revolution''~\cite{thhkat2015}.  Such a result has not been obtained hitherto
for a spatiotemporally chaotic NESS of a deterministic PDE. 

\twocolumngrid
\begin{widetext}
	
	\begin{figure}[htbp!] 
		\centering
		\includegraphics[width=1\textwidth,height = 0.995\textwidth]{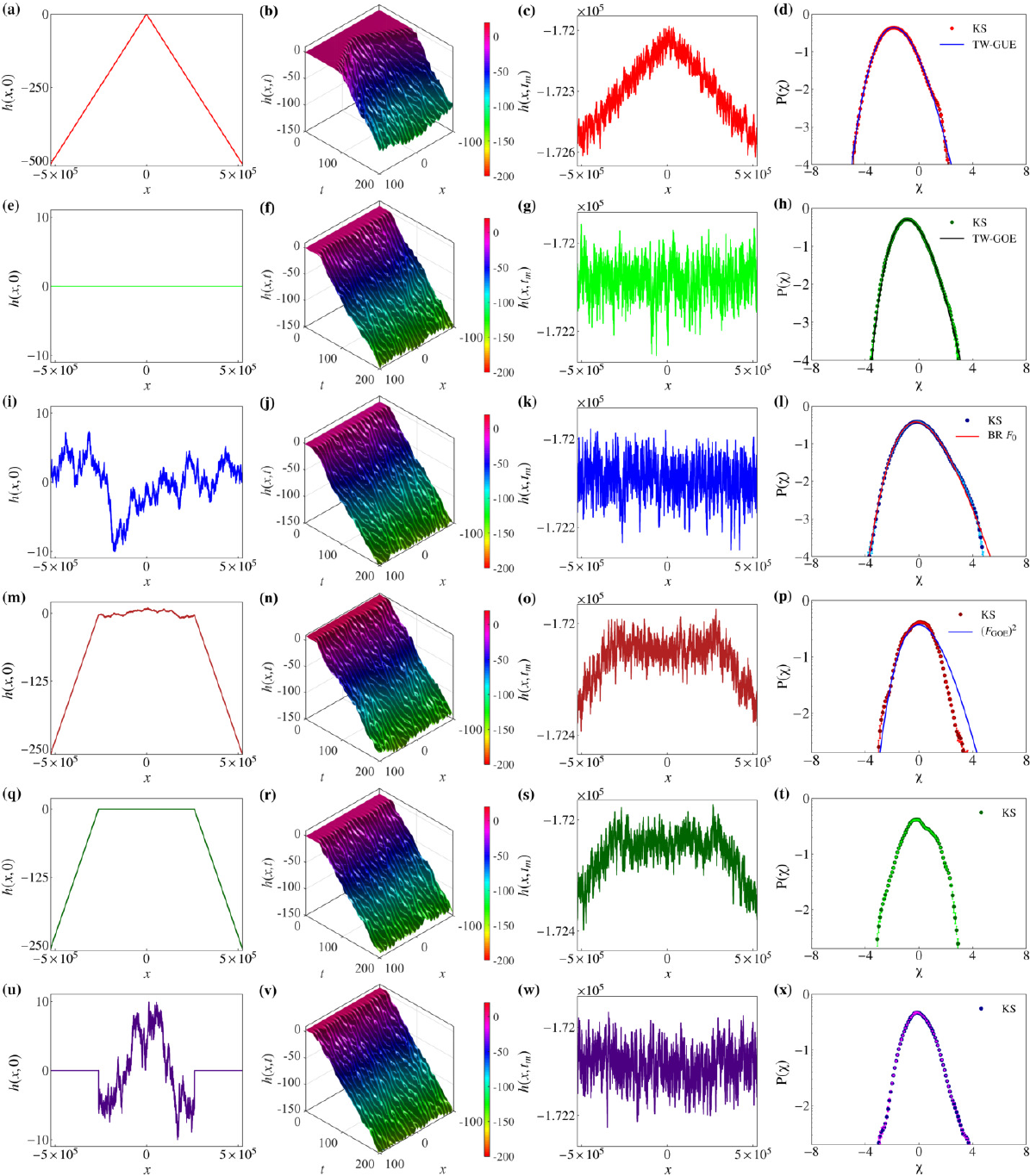}
		
		\caption{(Color online) Plots of $h(x,0)$ versus 
		$x \in \left[ -L/2, L/2 \right]$, with  $L=2^{20}$, for the six different 
		initial conditions, IC1, IC2, IC3, IC4, IC5, and IC6 in (a), (e), (i), 
		(m), (q), and (u), respectively. The short-time spatiotemporal evolution 
		of $h(x,t)$ is shown, in the interval $[-100,100]$, for each one of 
		IC1-IC6 in (b),(f),(j),(n),(r), and (v) (see the videos V1-V6 in the 
		Supplemental Material~\cite{supp}). The height profiles at time 
		$t_{m}=2\times10^5$ are plotted in (c), (g), (k), (o), (s), and
		(w) for IC1-IC6, respectively; and the plots (d), (h), (l), (p), (t),
		and (x) display corresponding limit distributions for $\upchi$ (see text)
		in the NESSs; and in (d), (h), (l), and (x) we plot 
		TW-GUE, TW-GOE, BR $F_{0}$, and $(F_{\text{GOE}})^2$ distributions to 
		compare them with data from our DNSs. The error bars on $\textrm{P}(\upchi)$ are smaller than the sizes of our symbols.} 						
		\label{fig:fig1}	
	\end{figure}
\end{widetext}

\begin{figure}[htbp!]
	\centering
	\includegraphics[width=1\linewidth]{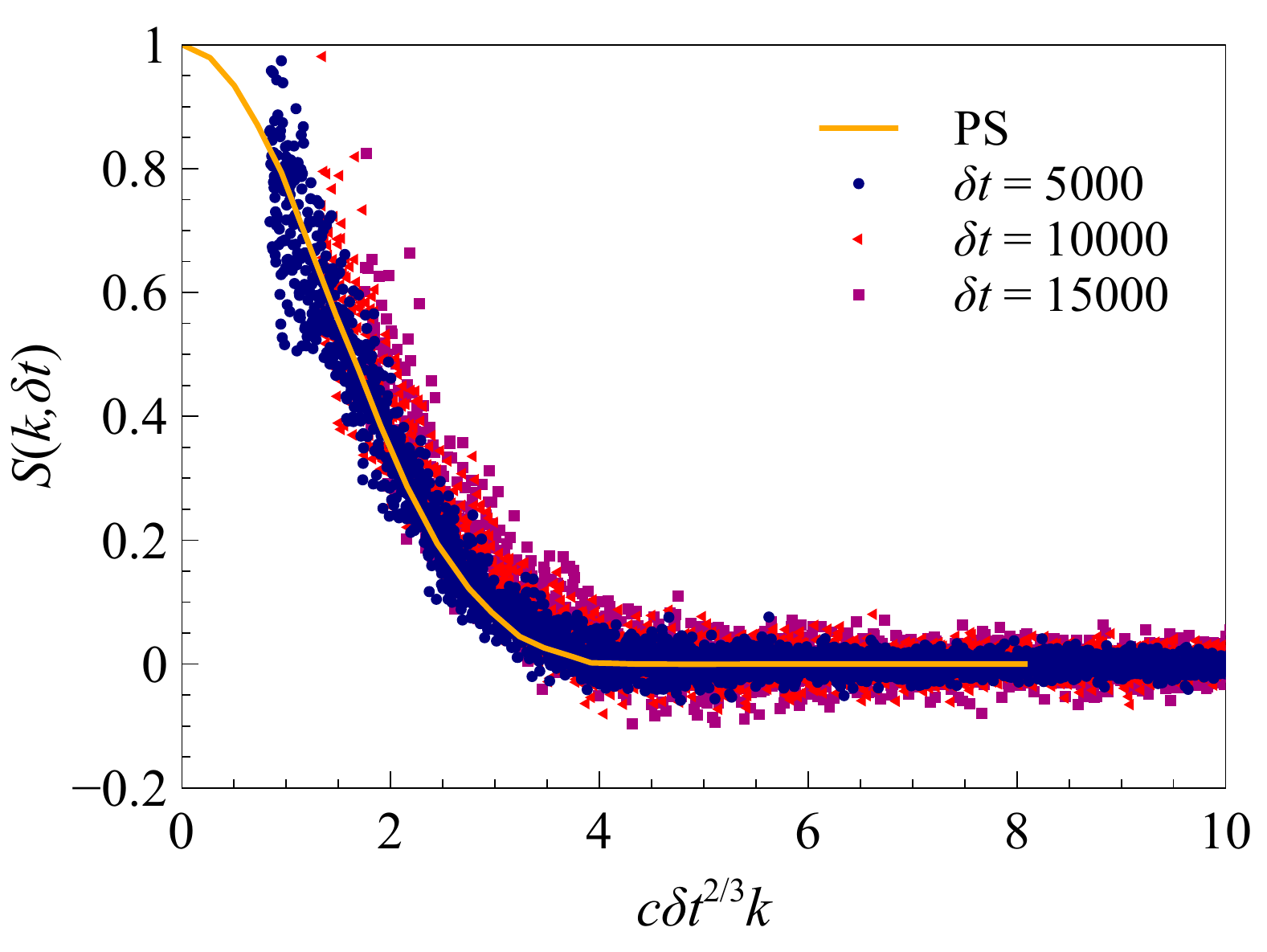}

	\caption{(Color online) Log-log plot of the scaling form of the Fourier
	transform of the two-point time-dependent correlation function $S(k,
	\delta t)$ versus $c \delta t^{2/3} k$, with the nonuniversal $c=1.6$, computed for IC3 (see
	Fig.~\ref{fig:fig1}(i)). We plot $S(k,
	\delta t)$ for three different values of $\delta t$; we also show, for
	comparison, the theoretical result (orange curve PS) obtained by
	Pr\"{a}hofer and Spohn~\cite{ps2004} for the 1D KPZ equation.}

\label{fig:fig2}	
\end{figure}

The KS PDE, which predates the KPZ SPDE, is
\begin{equation}
\partial_{t}h(\mathbf{x},t) + \Delta h(\mathbf{x},t) + \Delta^{2}h(\mathbf{x},t) +\frac{1}{2}(\nabla h(\mathbf{x},t))^{2}=0, \label{eq:ks}
\end{equation} 
where $\nabla \equiv \partial/\partial \mathbf{x}$, $\partial_{t} \equiv
\partial/\partial t$, $\Delta \equiv \nabla^{2}$, and $h, \, \mathbf{x}$, and
$t$ have been scaled such that the linear system size $L$ is the only control
parameter. The dynamical and long-wavelength properties of the 1D KS PDE
have been explored via DNSs in
Refs.~\cite{hnz1986, sneppen1992, hayot1993, hyman1986, kevrekidis1990};
several mathematical results have been obtained in Refs.~\cite{collet1992,
jolly1990, conte1989}.

The 1D  KPZ SPDE is 
\begin{eqnarray}
	\partial_{t} h(x,t) &=& \nu \Delta h(x,t) +\frac{\lambda}{2} (\nabla h(x,t))^2 + \eta \ , \nonumber \\
	\langle \eta(x,t) \eta(x',t') \rangle &=& D \delta(x-x')\delta(t-t')\ , \label{eq:kpz}
\end{eqnarray}
where $\nu$, the diffusivity, and $\lambda$, the strength of the nonlinearity, are real 
parameters, and $\eta$ is a zero-mean Gaussian white noise, with variance $D$.

We solve the 1D KS PDE~\eqref{eq:ks}, with periodic boundary conditions on a
domain of size $L$, by using the pseudospectral
method~\cite{cq1981,chqz2006,trefethen2000} and the $2/3$ dealiasing rule.  For
time marching we use the fourth-order, exponential time-differencing
Runge-Kutta scheme ETDRK4~\cite{kassam2005,cox2002}. For reliable statistics,
it is important to carry out long simulations with large values of $L$; we
report results with $L=2^{20}$, by far the highest spatial resolution that has
been used for a DNS of the 1D KS PDE~\eqref{eq:ks}; for this we have developed
a CUDA C code that runs very efficiently on a GPU cluster with NVIDIA Tesla K80
accelerators.

From our DNSs we compute $h(x,t)$ for six different kinds of initial
conditions, IC1-IC6, which we depict by plots of $h(x,0)$ versus $x$ in
Figs.~\ref{fig:fig1} (a), (e), (i), (m), (q), and (u); we show the short-time
spatiotemporal evolution of $h(x,t)$, in the interval $x \in [-100,100]$, in
Figs.~\ref{fig:fig1} (b), (f), (j), (n), (r), and (v) (see the videos V1-V6 in
the Supplemental Material~\cite{supp}). We choose these ICs to mimic the effect
of wedge, flat, stationary, wedge-to-stationary, wedge-to-flat, and
flat-to-stationary geometries in the ASEP model, which are listed in
Refs.~\cite{corwin2012,bfs2008,cfp2010} as initial conditions for six different
sub-classes of the 1D KPZ universality class.  Previous numerical
studies~\cite{hayot1993, sneppen1992} of the 1D KS PDE have shown that
two-point, equal-time height-field correlations show the scaling behaviors of
their 1D KPZ SPDE counterparts for times greater than a crossover time $t_{c}
\simeq 18700$ and lengths larger than the crossover size $L_{c}\simeq 3600$.
Therefore, we use a very large system size $L=2^{20}$ and very long simulation
times $t_{max} \geq 2\times10^5$ (see the Supplemental Material~\cite{supp}).

Our results for two-point height correlation functions are consistent with
those of earlier investigations~\cite{hayot1993, sneppen1992} of the
statistical properties of the spatiotemporally chaotic state of the 1D KS PDE:
We show, e.g., the equal-time compensated spectrum $k^2 E(k) =  \langle L
\tilde{h}(k,t) \tilde{h}^{*}(k,t) \rangle_{t} $, where $\langle \cdot
\rangle_{t}$ is the time average, $\tilde{h}(k,t)$ is the spatial Fourier
transform of $h(x,t)$, and $k$ is the wave number, in
Fig.~(\textcolor{blue}{1}) of the Supplemental Material \cite{supp}.  In
addition, we calculate the time-dependent, two-point correlation function $S(k,
\delta t) = \langle  k^2 \tilde{h}(k,t_{0}) \tilde{h}^{*}(k,t_{0} + \delta
t)\rangle_{t_{0}} $ in Fig.~\ref{fig:fig2}, for the IC3 initial condition. We
find that the imaginary part of $S(k, \delta t)$ fluctuates around zero and its
magnitude is much smaller than that of its real part, which we plot in
Fig.~\ref{fig:fig2}. Our data are consistent with the scaling form of $S(k,
\delta t)$ (orange curve in Fig.~\ref{fig:fig2}), which has been obtained
analytically by Pr\"ahofer and Spohn~\cite{ps2004} for the 1D KPZ SPDE; this
comparison of $S(k, \delta t)$ for the 1D KS and 1D KPZ equations has not been
made hitherto. 

The scaling properties of the interface width $w(L,t)$ distinguish different
universality classes in interface-growth models; 
\begin{equation}
	w(l,t) = \left( \langle [\Delta_{l} h(x,t)]^2 \rangle_{x, l}  \right)^{1/2}  ,
\end{equation}
with $\Delta_{l} h(x,t) = h(x,t) -h(x,0) - \langle h(x,t)-h(x,0) \rangle_{x,l}
$ and $\langle \cdot \rangle_{x,l}$ the spatial average over a region of
spatial extent $l$. For $t\gg1$ in the 1D KPZ equation, $w(L,t) \sim
t^{\upbeta}$.  Before crossover occurs in systems with $L > L_{c}$, the
exponent $\upbeta$ assumes the value $\upbeta_{\text{EW}}=1/4$, which is the
Edwards-Wilkinson (EW) result~\cite{edwards1982,thhzhang1995} for the linear
SPDE with $\lambda = 0$ in Eq.~(\ref{eq:kpz}); finally, $\upbeta$ assumes the
KPZ value $\upbeta_{\text{KPZ}}=1/3$ in the NESS (for $t>t_{c}$).  Moreover,
the growing KPZ surface involves the length scale $\mathcal{L}(t) \sim
t^{1/z}$, where the dynamic exponent $z = 3/2$; and the width $w(l,t)\sim
l^{\alpha}$, for $l \ll \mathcal{L}(t)$, with $\alpha=1/2$ \cite{takeuchi2011}.
We find from our DNSs of the 1D KS equation that these Family-Vicsek
scaling~\cite{fv1985} forms are indeed satisfied as we show in
Figs.~\ref{fig:fig3} (a), (c), and (e) for IC1-IC3 (see the Supplemental
Material~\cite{supp} for IC4-IC6). 


	\begin{figure}[htbp!] 
		\centering
		\includegraphics[width=1\linewidth]{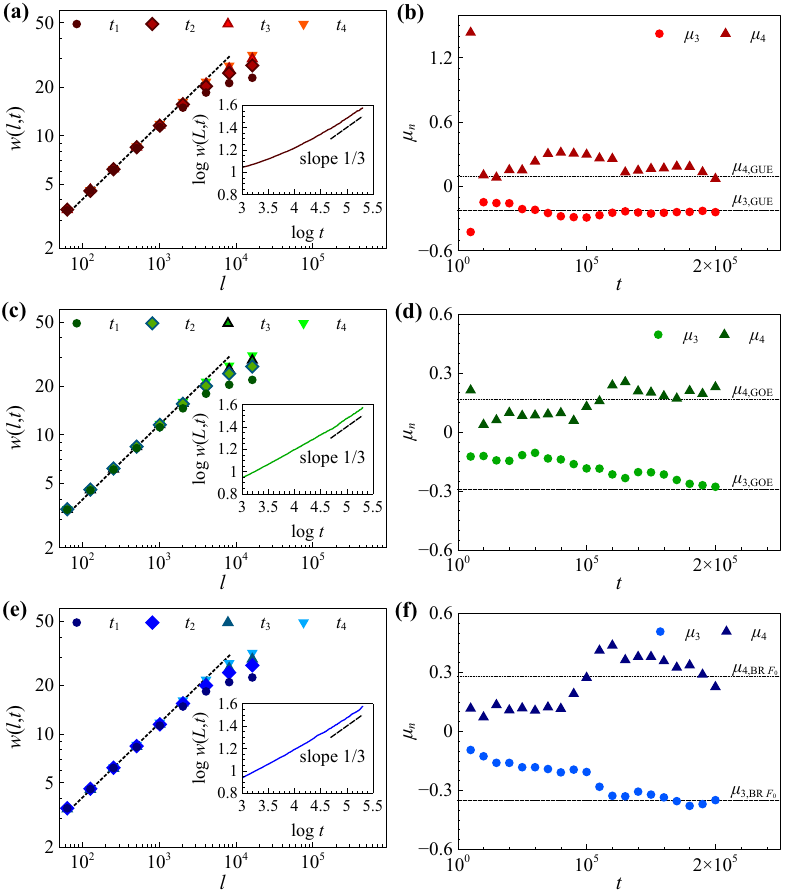}

		\caption{(Color online) Family-Vicsek scaling~\cite{fv1985}:
		(a), (c), and (e) show, for IC1-IC3, respectively, plots
		of $w(l,t)$ versus $l$, for $l \ll L$, and $w(L,t)$ versus $t$ (in the
		insets); $t_{1} = 5 \times 10^{4}$, $t_{2} =  10^{5}$, $t_{3} = 1.5
		\times 10^{5}$, and $t_{4} = 2 \times 10^{5}$. The dotted lines are log-log fits for $w(l,t)=A l^{\alpha}$, with $\alpha= 0.46 \pm 0.07$ for IC1-IC3. In (b), (d), and (f) we plot, 
		for IC1-IC3, respectively, the skewness $\mu_{3}$ and the kurtosis $\mu_{4}$ 
		(see text) versus the time $t$; black lines indicate their large-$t$ values for TW-GUE, TW-GOE, and BR 
		$F_{0}$ PDFs in (b), (d), and (f). (See the Supplemental Material~\cite{supp} for similar plots  for IC4-IC6.)}
		\label{fig:fig3}
	\end{figure}


We define 
\begin{equation}
	\mu_{n} = 	\langle  \left(
				\Delta_{L} h(x,t) \right)^{n}   \rangle / \langle  \left(  \Delta_{L} h(x,t)
				\right)^{2}   \rangle^{n/2} - 3 \delta_{n,4};
\end{equation}

for $n=3$ ($n=4$), $\mu_{n}$ is the skewness (kurtosis); we plot $\mu_{3}$ and
$\mu_{4}$ versus time $t$ in the right panel of Fig.~\ref{fig:fig3}; for each
initial condition, IC1-IC6, we average these quantities for $100$ surfaces,
over a time interval of $10^{4}$, and five independent DNS runs; i.e., our
overall sample size is $\simeq 5\times 10^8$ data points. [For our 1D KS,
$\mu_3 < 0$ because of the sign of the nonlinear term in Eq.~\eqref{eq:ks};
we ignore the sign of $\mu_{3}$ for it can be reversed by the transformation
$h(x,t) \rightarrow -h(x,t)$.] In addition, we calculate the probability
distribution function (PDF) $\textrm{P}(\upchi)$ of the shifted and rescaled
fluctuations, namely, $\upchi = (h(x,t) - v_{\infty} t)/(\Gamma t)^{1/3}$, when
both $\mu_{3}$ and $\mu_{4}$ are close to their standard values for the
relevant TW or BR $F_{0}$ PDFs; for IC2, e.g., we compute  $\textrm{P}(\upchi)$
when we have $\mu_{3} \simeq 0.27$ and $\mu_{4} \simeq 0.19$, which are close
to the standard values $\mu_{3,\text{GOE}} \simeq 0.29$ and $\mu_{4,\text{GOE}}
\simeq 0.16$, respectively. 

\begin{figure}[htbp!] 
	\centering
	\includegraphics[width=1\linewidth]{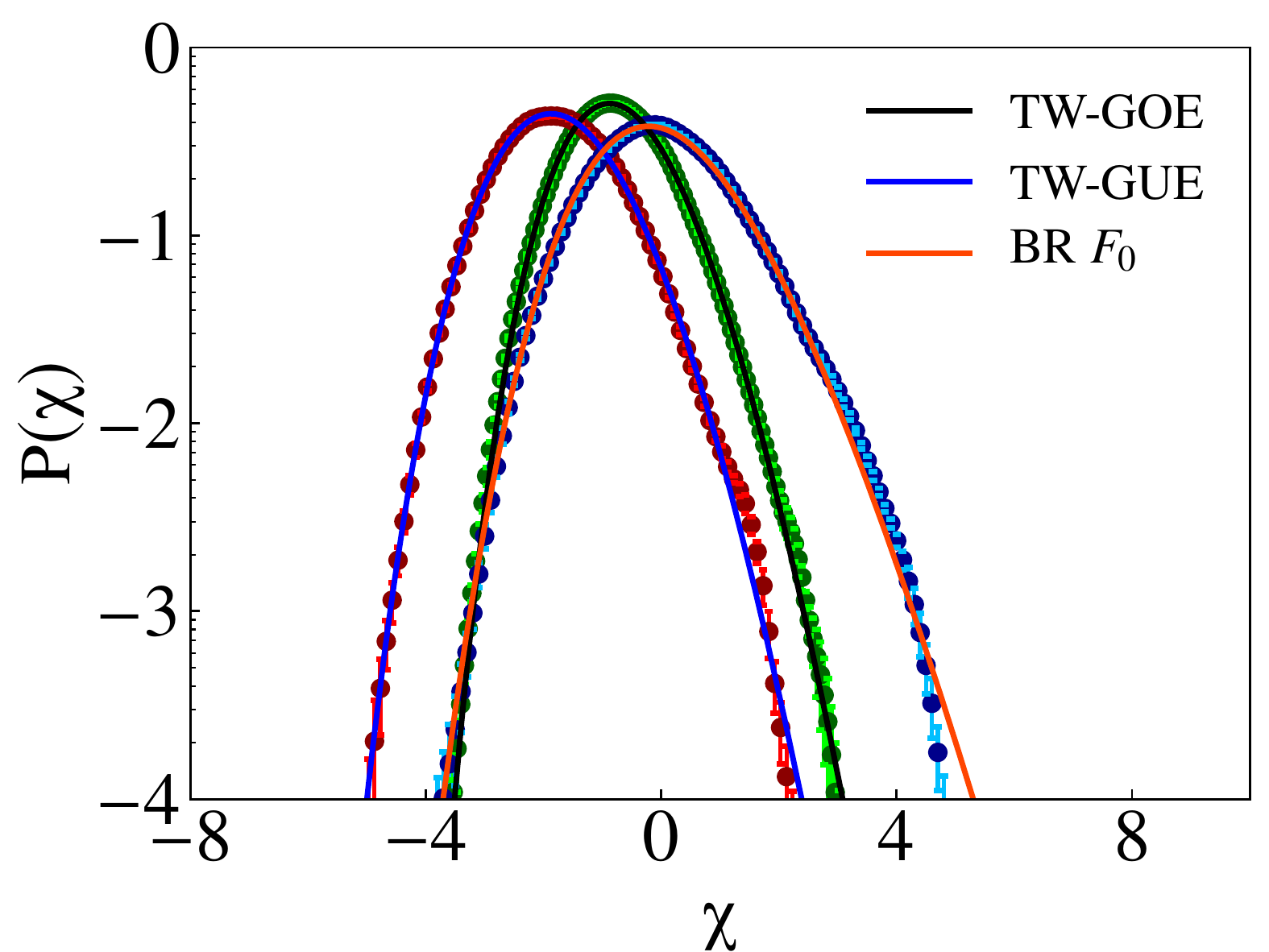}
	
	\caption{(Color online) Semilog plots of the PDFs $\textrm{P}(\upchi)$ from our DNSs for 
IC1, IC2, and IC3; we compare these with the Tracy-Widom distributions, TW-GUE and 
TW-GOE, and the Baik-Rains distributions (BR $F_{0}$).}
	\label{fig:fig4}
\end{figure}

For IC1, IC2, IC3, and IC4 we compare, on semilog plots, the PDFs with TW-GUE,
TW-GOE, BR $F_{0}$, and $(F_{\text{GOE}})^2$ \cite{corwin2012}
in Figs.~\ref{fig:fig1} (d), (h), (l), and (p), respectively. For ease of
comparison, we show in Fig.~\ref{fig:fig4} that the PDFs we obtain from our
DNSs of the  1D KS Eq.~\eqref{eq:ks} are very close to the TW-GUE, TW-GOE, and
BR $F_{0}$ PDFs over \textit{at least three orders of magnitude}. Stricly
speaking, we must collect data only from those two points ($x=L/4, 3L/4$) at
which the two different type of height profiles meet in cases IC4, IC5 and IC6.
However, this leads to inadequate statistics. Therefore, the PDFs of $\upchi$
for IC4-6, which we show  in Figs.~\ref{fig:fig1} (t) and (x), have been
computed by using data from the regions $ \left[ 7L/32, 9L/32\right]$ and
$\left[ 23L/32, 25L/32 \right]$; we see that this averaging procedure already
leads to PDFs (Figs.~\ref{fig:fig1} (p), (t) and (x)) that are distinctly
different from TW-GUE, TW-GOE, and BR $F_{0}$ distributions.

The TW distributions, for IC1 and IC2 initial conditions in the 1D KPZ
equation, have been studied in the context of $N\times N$ GOE ($\beta = 1$) and
GUE ($\beta = 2$) random matrices.  The largest eigenvalue (after scaling with
$N$) $\Lambda$ of such random matrices is
\begin{equation}
	\Lambda = \sqrt{2} +\frac{1}{\sqrt{2}} N^{-2/3} \upchi_{\beta} \ , \label{eq:evmax}
\end{equation}
where $ \upchi_{\beta}$ has the PDF~\cite{satya2014}
\begin{equation}
	\text{P}( \Lambda, N ) \approx  \begin{cases}
	\exp[-  \beta N^2 \phi_{-}(\Lambda)] ,   & \Lambda < \sqrt{2}, |\Lambda -\sqrt{2}| \sim \mathcal{O}(1),\\
	\sqrt{2} N^{2/3}\textrm{P}_{\textrm{TW},\beta}(\upchi_{\beta}) ,  \ & |\Lambda-\sqrt{2}| \sim \mathcal{O}(N^{-2/3}), \\
	\exp[- \beta N \phi_{+}(\Lambda)] ,   & \Lambda > \sqrt{2}, |\Lambda -\sqrt{2}| \sim \mathcal{O}(1),
	\end{cases} 
\end{equation}
$\textrm{P}_{\textrm{TW},\beta}(\upchi_{\beta})$ denotes TW distributions,
and the right and left large-deviation functions (LDFs) $\phi_{+}(\Lambda)$ and
$\phi_{-}(\Lambda)$, respectively, display the following asymptotic behaviors:
\begin{equation}
	\begin{aligned}
	\phi_{-}(\Lambda) &\approx  \frac{1}{6 \sqrt{2}} ( \sqrt{2} -\Lambda)^{3}   \ ,   \quad \Lambda \rightarrow  - \infty; \\
	\phi_{+}(\Lambda)	&\approx  \frac{2^{7/4}}{3} (\Lambda -\sqrt{2})^{3/2}  \ ,  \quad \Lambda \rightarrow  + \infty .
	\end{aligned}
	\label{eq:ldf}
\end{equation}  
The LDFs, which yield the probabilities of atypically large fluctuations, match
smoothly with the tails of $\textrm{P}_{\textrm{TW},\beta}(\upchi_{\beta})$.
Because of different behaviors of the tails of $\textrm{P}(\Lambda , N )$, a
third-order transition~\cite{satya2014} can be associated with $\Lambda$ at
$\Lambda_{c}= \sqrt{2}$ by defining the \emph{free energy} $ \propto \ln
F_{\beta}(\Lambda,N)$, $F_{\beta}(\Lambda,N)$ being the cumulative density
function (CDF) for $\Lambda$, for we have~\cite{satya2014}
\begin{equation}
	\lim_{N \rightarrow \infty} - \frac{1}{N^2} \ln F_{\beta}(\Lambda,N) = \begin{cases} 	\phi_{-}(\Lambda),  &  \Lambda < \sqrt{2}, \\
	0, & \Lambda >\sqrt{2}.
	\end{cases}
\end{equation}  
Similarly, we define, for the KS initial conditions IC1 and IC2, the
free-energy function $\mathcal{F}(\overline{h})$, for $t, L \rightarrow \infty
$, as follows:
\begin{equation}
	\mathcal{F}(\overline{h}) =  \lim_{t, L \rightarrow \infty} - \frac{1}{t^{2}} \mathrm{ln} \,  F(\upchi,t) \ ,   \label{eq:freenergy}	 
\end{equation}
where $\overline{h}= h(x,t)/t$ and $F(\upchi, t)$ is the CDF for $\upchi$ at
time $t$. Therefore, for IC1 and IC2, we should obtain a third-order phase
transition for $\overline{h}$ at the critical value
$\overline{h}_{c}=v_{\infty}$; an explicit demonstration requires much better
statistics for $\textrm{P}(\upchi)$ than is possible with our DNS.

We have shown, by extensive pseudospectral DNSs of the 1D KS deterministic PDE,
that the statistical properties of its spatiotemporally chaotic NESS are in the
1D KPZ universality class. This is not limited, merely, to the power-law forms
of simple correlation functions and the width of the interface.  It includes,
in addition, (a) the complete scaling form for the two-point time-dependent
correlation function $S(k, \delta t)$ (Fig.~\ref{fig:fig2}), (b) the skewness
and kurtosis shown in Fig.~\ref{fig:fig2}, and (c) most important of all, the
unversal limit distributions in Fig.~\ref{fig:fig1}, obtained in ``the $2^{nd}$
KPZ Revolution''~\cite{thhkat2015}. Such results have not been obtained
hitherto for a spatiotemporally chaotic NESS of any deterministic PDE. We
conjecture that similar conclusions should ensue for the phase-chaos regime of
the 1D Complex-Ginzburg-Landau equation~\cite{grinstein1996}. Such studies are
also being pursued for the 1D Calogero-Moser model~\cite{aka2019}. 

\begin{acknowledgments}
We thank Jaya Kumar Alageshan,  R. Basu, M. Brachet, P. Ferrari, T. Imamura, K.
Khanin, and K. A. Takeuchi for discussions and the National Mathematics
Initiative (NMI), DST, UGC, and CSIR (India) for support.
\end{acknowledgments}

\bibliographystyle{apsrev4-1}
\bibliography{KSE.bib}

\end{document}


\title{Supplemental Material: The one-dimensional Kardar-Parisi-Zhang and Kuramoto-Sivashinsky 
universality class: limit distributions}
\author{Dipankar Roy\textsuperscript{*}}
\address{Department of Mathematics, Indian Institute of Science,\\
	Bangalore - 560012, India.\footnote{\textsuperscript{*}dipankarroy@iisc.ac.in}}
\author{Rahul Pandit\textsuperscript{\dag}}
\address{Centre for Condensed Matter Theory, \\
	Department of Physics,
	Indian Institute of Science,\\
	Bangalore - 560012, India.
	\footnote{\textsuperscript{\dag}rahul@iisc.ac.in; also at Jawaharlal Nehru Centre for Advanced Scientific Research, Jakkur, Bangalore 560 064.}}

\date{\today}
\maketitle

This Supplemental Material contains details of the calculations that we have 
presented in the main part of this paper.

\section{The compensated spectrum}
\begin{figure}[htbp!]
	\centering
	\includegraphics[width=0.5\linewidth]{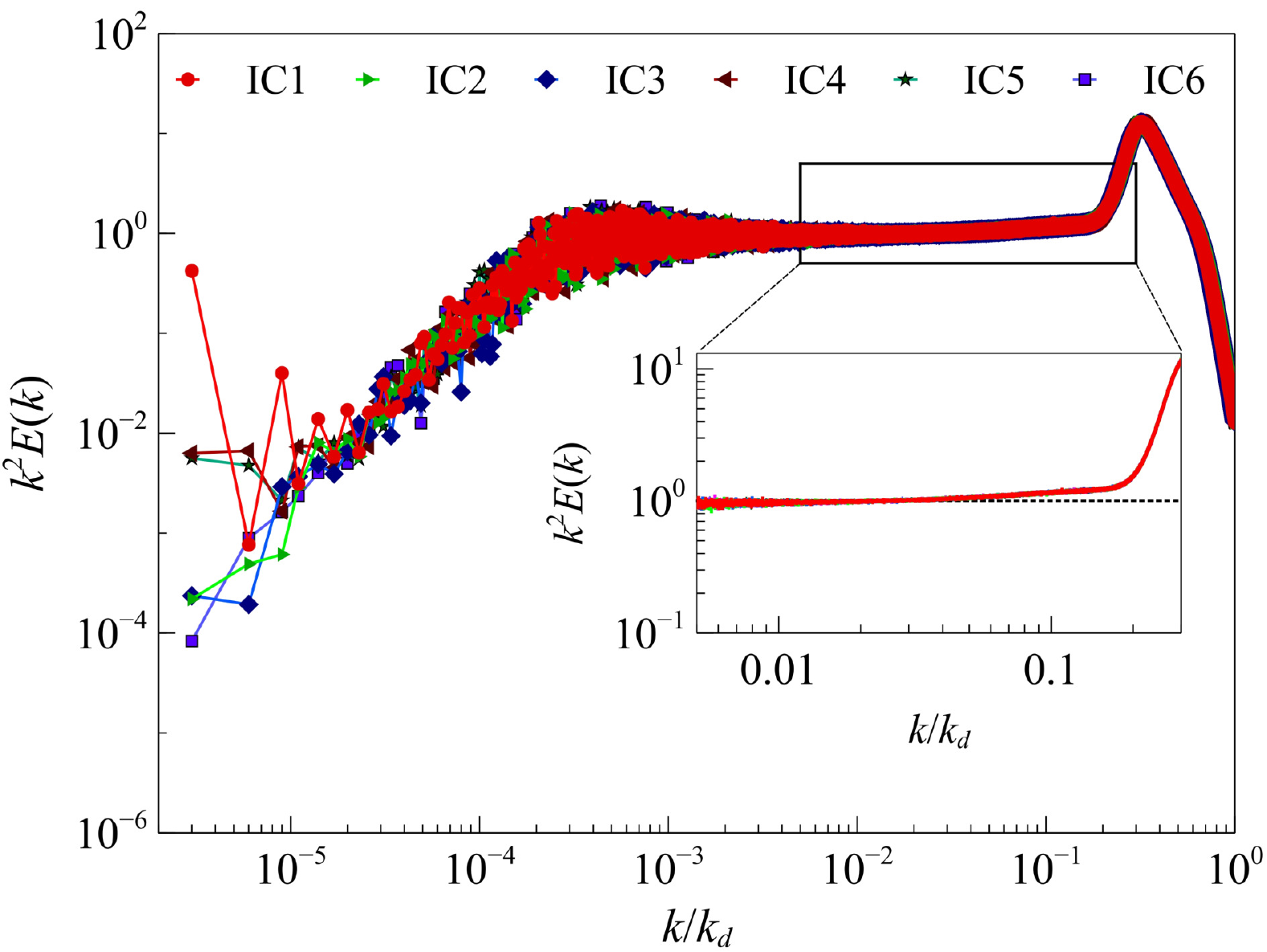}
	\caption{(Color online) Log-log plots of the compensated spectrum $k^2 E(k)$ versus $k/k_{d}$ for the six different initial conditions IC1-IC6 (see Fig.~(\textcolor{blue}{1}) of the main text). We zoom into the region $\delta k/k_{d}=[0.005, 0.3]$, where the curves appear flat, and show, in the inset, how our data compare with the line $k^2 E(k) = 1$. Here, $k_{d}= \pi \lfloor L/3\rfloor/L$ is the value of the maximum wave number after dealiasing and the system size $L=2^{20}$. }
	\label{fig:spectrum}	
\end{figure}

We show the equal-time
compensated spectrum $k^2 E(k) =  \langle L \tilde{h}(k,t) \tilde{h}^{*}(k,t)
\rangle_{t} $, where $\langle \cdot \rangle_{t}$ is the time average and
$\tilde{h}(k,t)$ is the spatial Fourier transform of $h(x,t)$ and $k$ is the wave number, in Fig.~\ref{fig:spectrum}.
The compensated spectrum $k^2 E(k)$ that we present in Fig.~\ref{fig:spectrum} covers a much larger range of wave numbers than in earlier simulations \cite{hayot1993,sneppen1992}.

\section{Computation of the parameters $v_\infty$ , $\Gamma$, and $\upbeta$}

\begin{figure}[htbp!]
	\centering
	\begin{subfigure}{0.475\linewidth}
		\includegraphics[width=1\linewidth]{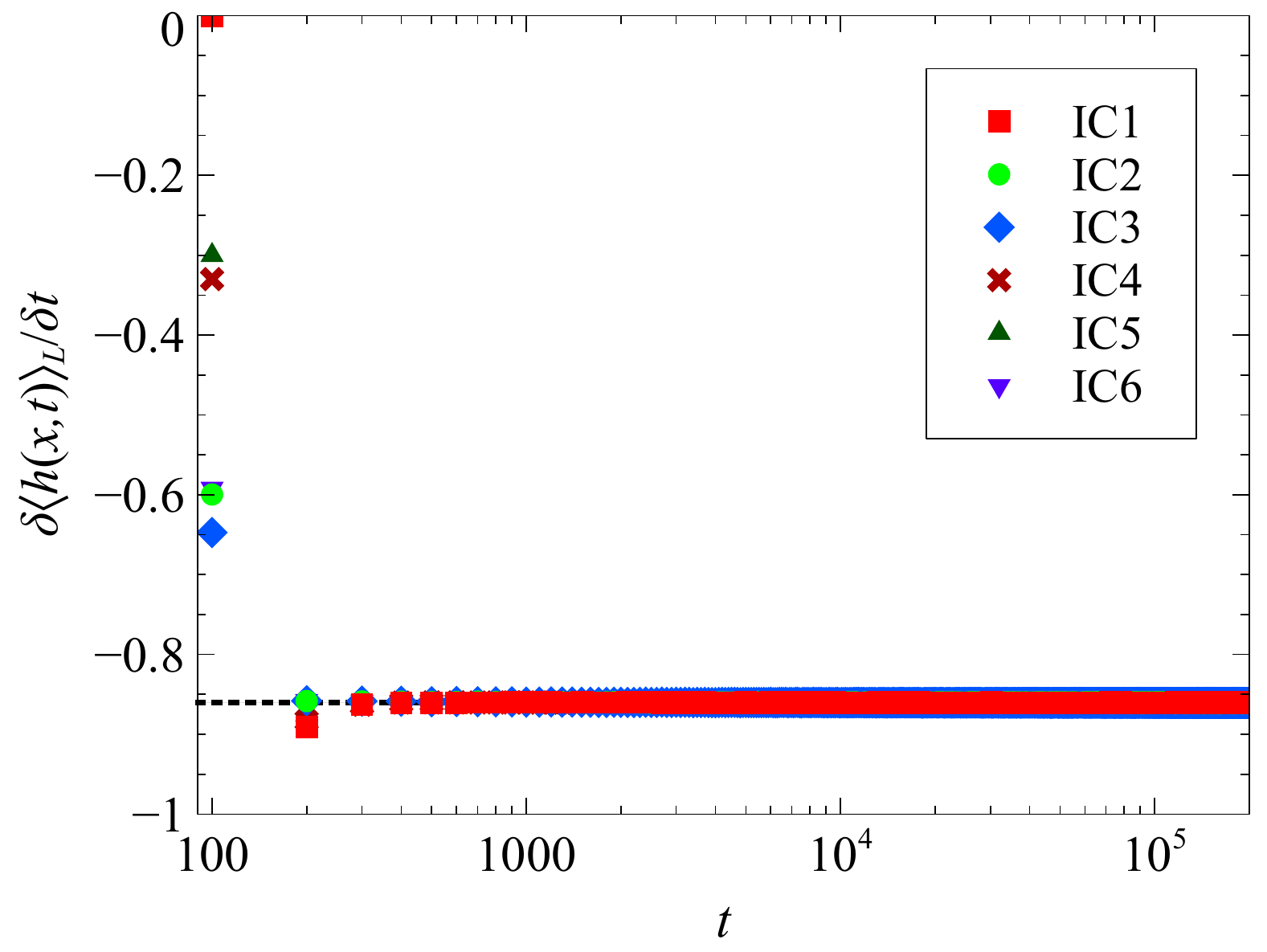}
		\caption*{(a)}
	\end{subfigure}%
	\begin{subfigure}{0.475\linewidth}
		\includegraphics[width=1\linewidth]{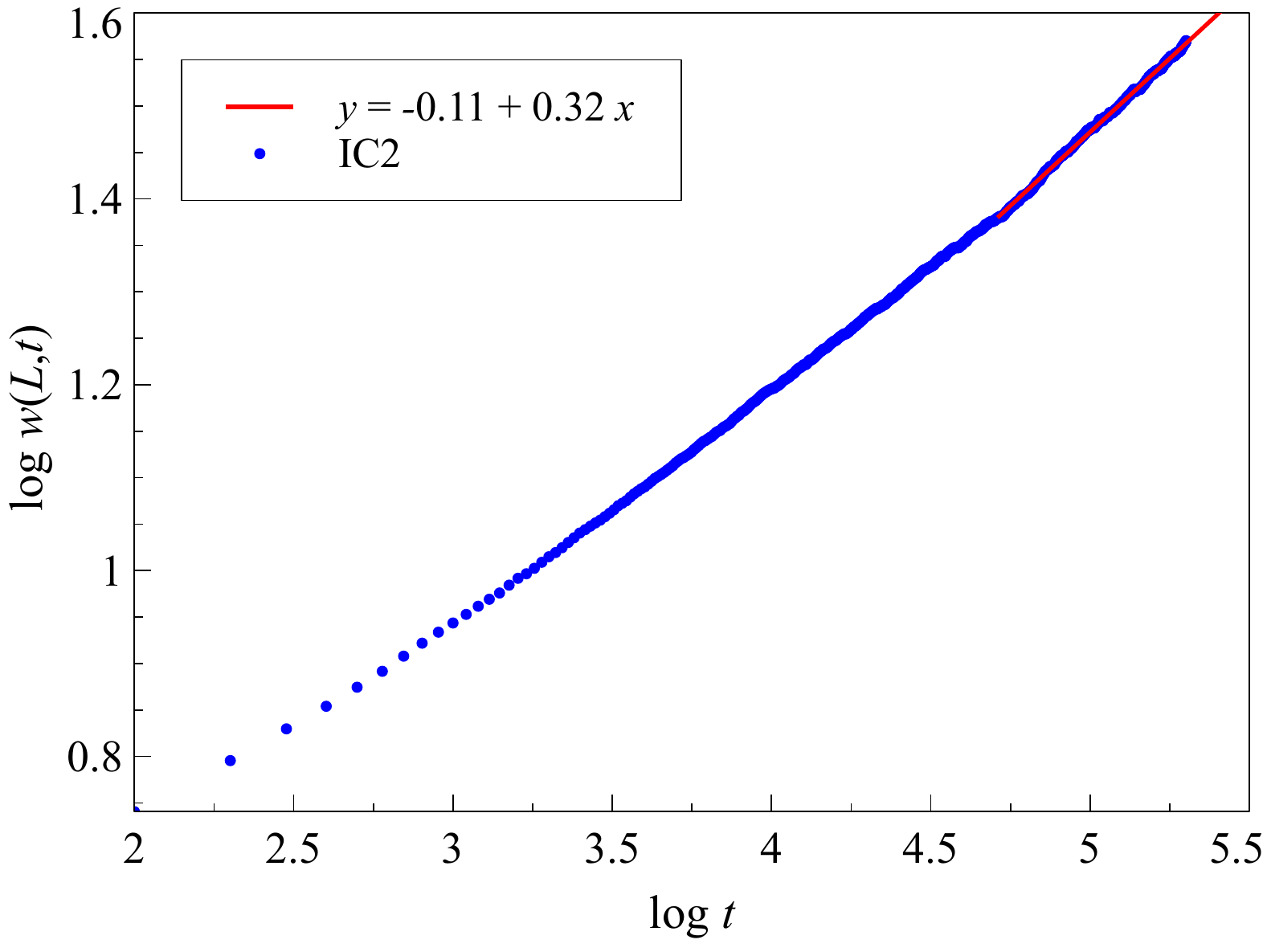}
		\caption*{(b)}
	\end{subfigure}%
	\caption{(Color online) We plot $ \langle \delta h(x,t) \rangle_{L} / \delta t$ versus $t$ in (a). In (b), we display $\log w(L,t)$ versus $\log t$.}
	\label{fig:fig1}	
\end{figure}

We compute the model-dependent parameters $v_{\infty}$ and $\Gamma$
(see Eq.~(\textcolor{blue}{1}) in the main paper) from our DNS data as follows. By choosing two
Kuramoto-Sivashinsky (KS) surfaces at two different times with $ \delta t
=100$, we compute $ \delta \langle h(x,t) \rangle_{L} / \delta t$, where
$\langle \cdot \rangle_{L}$ is the spatial average over our simulation domain;
we plot it versus time $t$ in Fig.~\ref{fig:fig1}~(a); the $t \rightarrow
\infty$ limit yields $v_{\infty} \simeq -0.86$. 

The exponent $\upbeta$ appears in the Family-Vicsek scaling form:
\begin{equation}
	w(L,t) \sim t^{\upbeta}, \text{ for } t \rightarrow \infty,
\end{equation}
where $w(L,t)$ is the width (see the main paper). To compute $\upbeta$ we plot
$\log w(L,t)$ for IC2 (the flat initial condition) against $\log t$ in
Fig.~\ref{fig:fig1}~(b); a linear fit yields the exponent $\upbeta \simeq
0.32$, which is close to the KPZ value $\upbeta_{\text{KPZ}} = 1/3$. 

\begin{figure}[htbp!]
	\centering
		\includegraphics[width=0.5\linewidth]{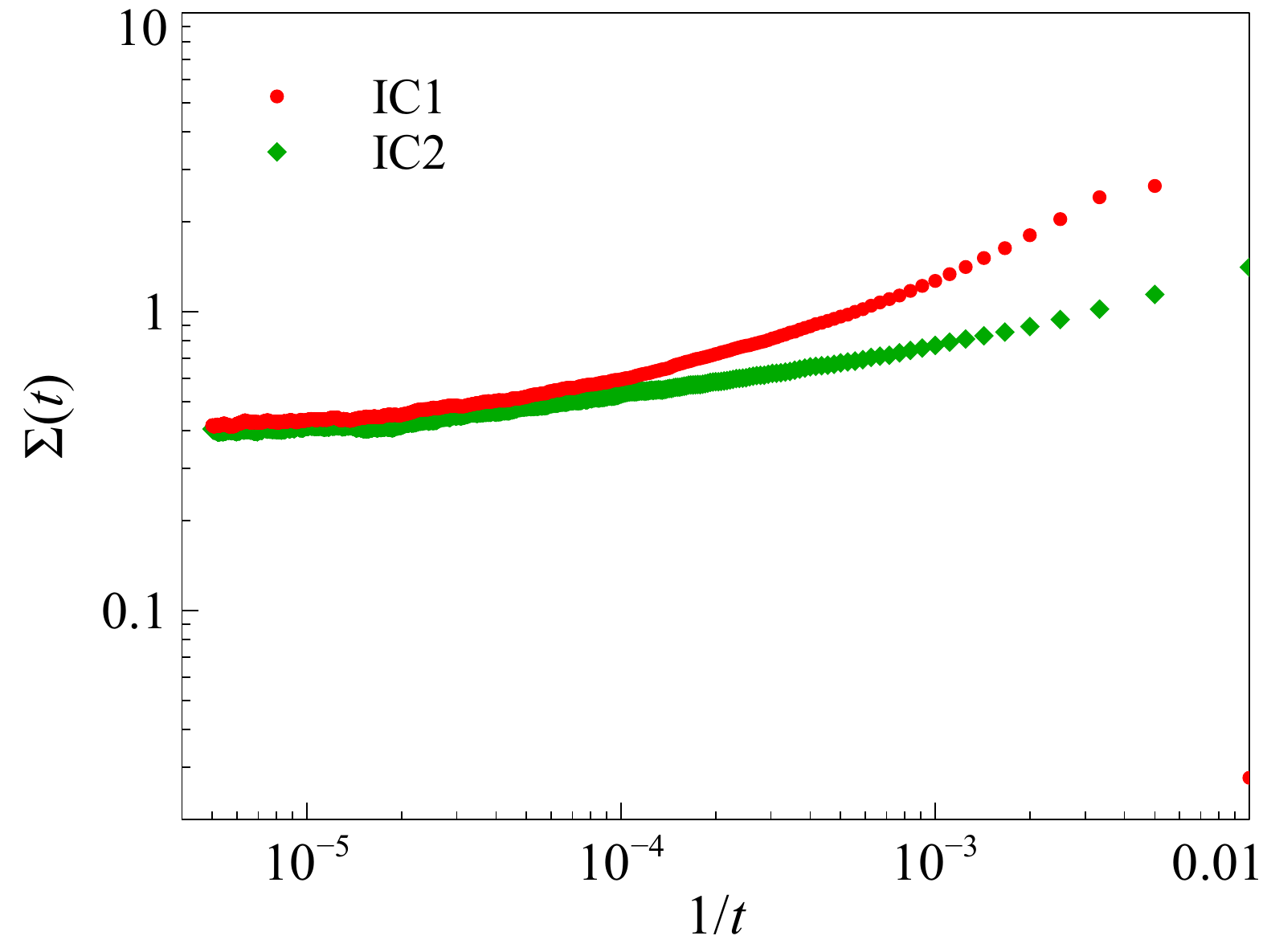}
		\caption{(Color online) Log-log plots of $\Sigma(t)$ versus $t$ for IC1 and IC2.}
		\label{fig:gamma}	
\end{figure}

In order to find the constant $\Gamma$, we compute $\Sigma(t)$, the variance
of $ (h(x,t) - v_{\infty} t) / t^{\upbeta_{\text{KPZ}}}$, and plot
it versus $t$ (see the log-log plot in Fig.~\ref{fig:gamma}). For $t \gg 1$, we
have $\Sigma(t) \rightarrow \Gamma^{2/3} \textrm{Var}(\upchi_{\beta})$, where 
$\textrm{Var}(\upchi_{\beta})$ is the variance of the random variable $\upchi_{\beta}$ with
$\beta = 1$ for IC2 and $\beta=2$ for IC1. Given that the variances of the PDFs of
$\upchi_{1}$ and $\upchi_{2}$ are, respectively, $\simeq 0.638$ and $\simeq0.813$ 
(see Ref.~\cite{prahofer2000}), we compute $\Gamma \simeq 0.358 $ and $\Gamma \simeq
0.496$ for IC1 and IC2, respectively. 

\section{Family-Vicsek scaling and the skewness and kurtosis for IC4-IC6}

\begin{figure}[htbp!] 
	\centering
	\includegraphics[width=1\linewidth]{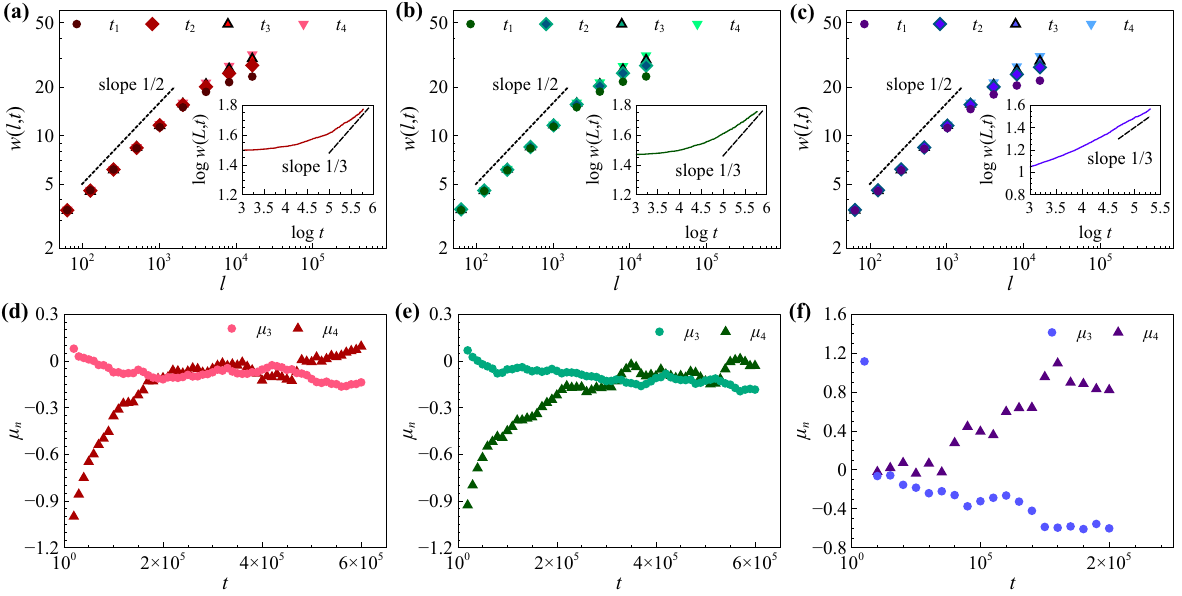}
	
	\caption{(Color online) Plots of Family-Vicsek scaling in (a)-(c), and the skewness and kurtosis in (d)-(f) for IC4-IC6, respectively.}
	\label{fig:fig2}
\end{figure}

In Figs.~(\ref{fig:fig2})~(a)-(c) we show Family-Vicsek scaling for the initial
conditions IC4-IC6. In Figs.~(\ref{fig:fig2})~(d)-(f) we plot the skewness and
kurtosis for these initial conditions. Stricly speaking, we must collect 
data only from those two points ($x=L/4, 3L/4$) at which the two different type of height profiles meet in cases IC4, IC5, and IC6. However, this leads to inadequate statistics. Therefore, the skewness and kurtosis are computed by using data from the regions $ \left[7L/32, 9L/32\right] $ and $ \left[ 23L/32, 25L/32 \right] $.

\section{Simulation details \label{supp_sec:simulation}}

We have used the exponential time-differencing fourth-order
Runge-Kutta method (ETDRK4) \cite{cox2002,kassam2005} for time marching in our direct numerical simulation (DNS) for the 1D KS equation. The whole simulation is programmed in CUDA C, by using the in-built
\textit{Fast Fourier Transform} in CUDA to switch back and forth between
Fourier and real space in our pseudospectral DNS. Moreover, the $2/3$ dealiasing rule is incorporated to
avoid aliasing errors. The parameters for our DNSs are given in
Table~\ref{table:dnsparams}.

\begin{table}[htbp!]
	\renewcommand*{\arraystretch}{1.6}
	\begin{tabular}{|c|c|c|c|c|}
		\hline 
		$L$ &  $N$ & $\delta x$ & $\delta t$ &  $t_{max}$ \\  
		\hline 
		$2^{20}$ & $2^{20}$ & 1 & 0.01 &  $ 2-6 \times 10^5$ \\ 
		\hline 
	\end{tabular} 
	\vspace{0.2cm}
	\caption{DNS parameters: $L$ is the system size, $N$ is the number of collocation points, $\delta x  = L/N$, $\delta t$ is the time step, and $t_{max}$ is the maximum time for which we run our DNS.}
	\label{table:dnsparams}
\end{table}

The evolution of $h(x,t)$ for the six initial conditions IC1-IC6 is captured in
the videos that are available at the URLs provided below:
 
\begin{itemize}
	\item IC1 : {\scriptsize \url{https://drive.google.com/open?id=1CPqNxda1GbntAmgHDydg3xzDRsqycxrr}}
	
	\item IC2 : {\scriptsize \url{https://drive.google.com/open?id=1iL154onInbzeCTzjmRJgZ-osgf6TuhcN}}
	
	\item IC3 : {\scriptsize \url{https://drive.google.com/open?id=1xRcXlFrETj1VqUYC5fYXeZ9s0oORw-V4}}
	
	\item IC4 : {\scriptsize \url{https://drive.google.com/open?id=1RULXEHa-gz4i8vGKNzH2Q54wQo4GTemw}}
	
	\item IC5 : {\scriptsize \url{https://drive.google.com/open?id=14rmSiAmzhBFSQ84HNj_9MOn7U_MWzzyC}}
	
	\item IC6 : {\scriptsize \url{https://drive.google.com/open?id=131uJ5mBO8DCifAUovsr4i3FHN4Zh7758}}
\end{itemize}

\bibliographystyle{plain}
\bibliography{Supplemental}